\documentclass{llncs}
\usepackage{graphicx}
\usepackage{amsmath}
\usepackage{color}
\usepackage{balance}
\usepackage{oz}
\usepackage[ruled,vlined,linesnumbered]{algorithm2e}
\usepackage{lscape}

\begin{document}

\title{Identifying Implicit Vulnerabilities through Personas as Goal Models}
\author{Shamal Faily \inst{1}\orcidID{0000-0002-2859-1143} \and Claudia Iacob \inst{2}\orcidID{0000-0002-1032-7598} \and Raian Ali \inst{3}\orcidID{0000-0002-5285-7829} \and Duncan Ki-Aries \inst{1}\orcidID{0000-0001-8114-2737}}
\authorrunning{Faily et al.}

\institute{Department of Computing \& Informatics, Bournemouth University, Poole, UK\\
\email{\{sfaily,dkiaries\}@bournemouth.ac.uk}
\and
School of Computing, University of Portsmouth, UK\\
\email{claudia.iacob@port.ac.uk}
\and
Hamid Bin Khalifa University, Doha, Qatar\\
\email{raali2@hbku.edu.qa}}

\maketitle

\begin{abstract}
When used in requirements processes and tools, personas have the potential to identify vulnerabilities resulting from misalignment between user expectations and system goals.  Typically, however, this potential is unfulfilled as personas and system goals are captured with different mindsets, by different teams, and for different purposes.  If personas are visualised as goal models, it may be easier for stakeholders to see implications of their goals being satisfied or denied, and designers to incorporate the creation and analysis of such models into the broader RE tool-chain.  This paper outlines a tool-supported approach for finding implicit vulnerabilities from user and system goals by reframing personas as social goal models.  We illustrate this approach with a case study where previously hidden vulnerabilities based on human behaviour were identified. 
\end{abstract}

\section{Introduction}
\label{sect:introduction}

Personas are fictional characters that represent archetypal users, and embody their needs and goals \cite{core14}.  Personas are the product of research with representative end-users, so designing for a single persona means designing for the user community he or she represents.  By facilitating design for one customer voice rather than many, personas have become a popular User Experience (UX) technique for eliciting and validating user requirements.

Personas can be a useful addition to requirements processes and tools when `building security in'.   If we identify that a persona experiences physical or cognitive burden while completing a task then its performance might not be as intended.  Steps might be omitted or the task altered to achieve an end more conducive to the persona's own goals, irrespective of whether or not the intent is malicious.  

Personas can inspire the identification of security vulnerabilities.  In practice, they usually do not.  Design processes prioritising agility provide little time for using personas for anything besides validating stakeholder value has been achieved.  Even if we assume UX and security engineers collaborate, personas are not always used in the ways envisaged by designers \cite{frie12}, while security engineers might primarily focus on requirements for security mechanisms.  Given their differing concerns and perspectives, problems may not be found even when these are indicated during the collection or analysis of user research data.

Personas, as user models, can be integrated into Security Requirements Engineering (RE) practices and tools, but they need to be built and presented differently.  This may make it easier for stakeholders to identify the security implications of user goals being satisfied or denied.  Goal models in languages like \emph{i*} \cite{yu97} and the Goal-oriented Requirements Language (GRL) \cite{amgh10} provide a foundation for this improved integration; they represent the intentions and rationale of social and technical actors, their inter-relations, and alternative strategies giving a space for variability accommodation, including that of user types.  Approaches like Secure Tropos \cite{mogi072} and STS-ml \cite{padg13} show how goal models can be used in the early stage of design to find vulnerabilities.  However, they are role-focused whereas people are expected to align to one or more ways to achieve predefined goals.

To integrate personas into Goal-oriented Security Requirements Engineering, we need to answer two research questions.  First, how can persona creation be leveraged to construct goal models (RQ1)?  Second, how can existing goal modelling approaches and RE tools, with minimal changes, be constructed to reveal \emph{implicit} vulnerabilities -- vulnerabilities that may be present when dependees fall short of their responsibility to deliver dependums \cite{lymy03} -- without burdening designers with additional conceptual knowledge (RQ2)?  User research and threat modelling can be time-consuming and cognitively intensive activities that might happen separately or in parallel before, during, or after other Requirements Engineering activities.  It is, therefore, necessary to loosely couple these goal models such that other design models can evolve orthogonally with minimum disruption to existing processes and tools.

In this paper, we present a tool-supported approach for finding implicit vulnerabilities by reframing personas as social goal models.  The remainder of this paper is structured as follows.  In Section \ref{sect:relatedwork}, we consider related work in social goal modelling and security, personas, and usable \& secure Requirements Engineering upon which our approach is based.  In Section \ref{sect:approach}, we present the processes and tool-support algorithms that underpin our approach before describing its application to an industrial control systems case study example in Section \ref{sect:results}.  We discuss the implications of our work and potential limitations in Section \ref{sect:discussion}, before concluding in Section \ref{sect:conclusion} by summarising the contributions of our work to date, and directions for future work.

\section{Related Work}
\label{sect:relatedwork}

\subsection{Finding vulnerabilities using social goal modelling}

Social goal modelling languages like \emph{i*} capture the modelling of dependencies, where a \emph{depender} actor depends on \emph{dependee} actor for some resource \emph{dependum}.  Actors become vulnerable when they rely on dependees for dependums.  Analysing chains of these dependencies can help us understand how vulnerable these actors are \cite{yu95}.  Moreover, when such models capture a socio-technical system of actors and resources, they can also highlight potential system vulnerabilities resulting from inconsistencies between an organisation's policies and working practices \cite{maza11}.

In previous work examining the use of social goal modelling to support Security Requirements Engineering, Liu et al. \cite{lymy03} considered how legitimate actors might use their intentions, capabilities and social relationships to attack the system, and how dependency relationships form the basis of exploitable vulnerabilities.  The idea of dependencies as implicit vulnerabilities was further elaborated by Giorgini et al. \cite{gmmz05}, who indicated that dependency relationships can also capture trust relationships where dependers believe dependees will not misuse a goal, task or resource (Trust of permission), or a trustee believes dependees will achieve a goal, execute a task, or deliver a resource (Trust of execution).

Elahi et al. \cite{elyu09} incorporated vulnerabilities into goal models to link knowledge about threats, vulnerabilities, and countermeasures to stakeholder goals and security requirements.  Vulnerabilities are considered as weaknesses in the structure of goals and activities of intentional agents, which can be propagated via decomposition and dependency links.  The introduction of vulnerabilities was added on the basis that including security and non-security elements on a single model makes models clearer and facilitates model discussion \cite{siop07}.  However, while this approach supports the \emph{specification} of vulnerabilities, it provides little support for \emph{eliciting} them.  This still requires a priori knowledge of potential system weaknesses or threat models that could take advantage of them.  Moreover, Moody et al. \cite{mohe09} found that the graphical complexity of \emph{i*} is several times greater than a human's standard limit for distinguishing alternatives.  As such, approaches that increase the complexity of the \emph{i*} language are likely to hinder rather than improve the understandability of social goal models, particularly for novices.

\subsection{Personas for security}

UX professionals have long used personas to bring user requirements to life, and there has been some been work within the Requirements Engineering community on using personas to add contextual variability to social goal models, e.g. \cite{rtwa18}.

The merits of using personas to explicitly elicit security requirements was identified by Faily et al. \cite{fafl106}, who showed how the use of personas could show the human impact of security to stakeholders who have never met user communities represented by personas.  In recent years, there has also been additional interest in the RE community on the use of personas to engage stakeholders when validating requirements \cite{clel13}, and how data used to construct personas can have some security value.  For example, Mead et al. \cite{mssh17} demonstrated how the text from personas built on assumptions about attackers (Personae Non Gratae) could be mined to identify potential threat models and identify gaps between a designer's and attacker's model of a system.  However, Mead et al. focuses on the identification of threats to a system rather than vulnerabilities that might arise from interactions between personas and the system. 

\subsection{IRIS and CAIRIS}

IRIS (Integrating Requirements and Information Security) is a process framework for designing usable and secure software \cite{fail18}.  The framework incorporates a methodology agnostic meta-model for usable and secure requirements engineering that supports the complementary use of different Security, Usability, and Requirements Engineering techniques.  Personas are integrated into this framework, which uses the KAOS language for modelling system goals \cite{lams09}, obstacles that obstruct the satisfaction of these goals, dependency associations between roles, and relationships between tasks, system goals, and the roles responsible for them.  The framework is complemented by CAIRIS (Computer-Aided Integration of Requirements and Information Security): a software platform for eliciting, specifying, automatically visualising, and validating secure and usable systems that is built on the IRIS meta-model.  By making explicit the links between different security, usability, and software models using IRIS, and providing tool-support for automating generating and validating these models, IRIS and CAIRIS can put one model in context with another.  For example, we recently demonstrated how data flow \emph{taint} could be identified in data flow diagrams within CAIRIS by putting these diagrams in context with other software and usability models \cite{fsss20}.

Previous work has shown that, if personas are constructed using qualitative data analysis, the results of this analysis can be framed as argumentation models \cite{fafl111}, and the elements of these models can be re-framed as goals and soft goals in social goal models \cite{failyre11}.  Not only does this make it possible to automatically generate goal models from argumentation models, some assurance is also provided for both the basis of user goals and the broader impact of satisfying these goals on other system elements.  Subsequent work has demonstrated how these concepts can lead to generation of elaborate GRL compatible goal models \cite{fafl142}.  However, a weakness of this approach is its reliance on additional tool-support (jUCMNav), and the limited support of traceability links between the goal modelling platform and its originating data should the GRL model evolve; such evolution is likely as different stakeholders make sense of this model.  Subsequent refinement of the jUCMNav model could lead to additional effort by analysts to ensure the goal model and its foundational CAIRIS models are synchronised.

\section{Approach}
\label{sect:approach}

\subsection{Conceptual model}

\begin{figure}[h!]
\centering
\includegraphics[width=\textwidth]{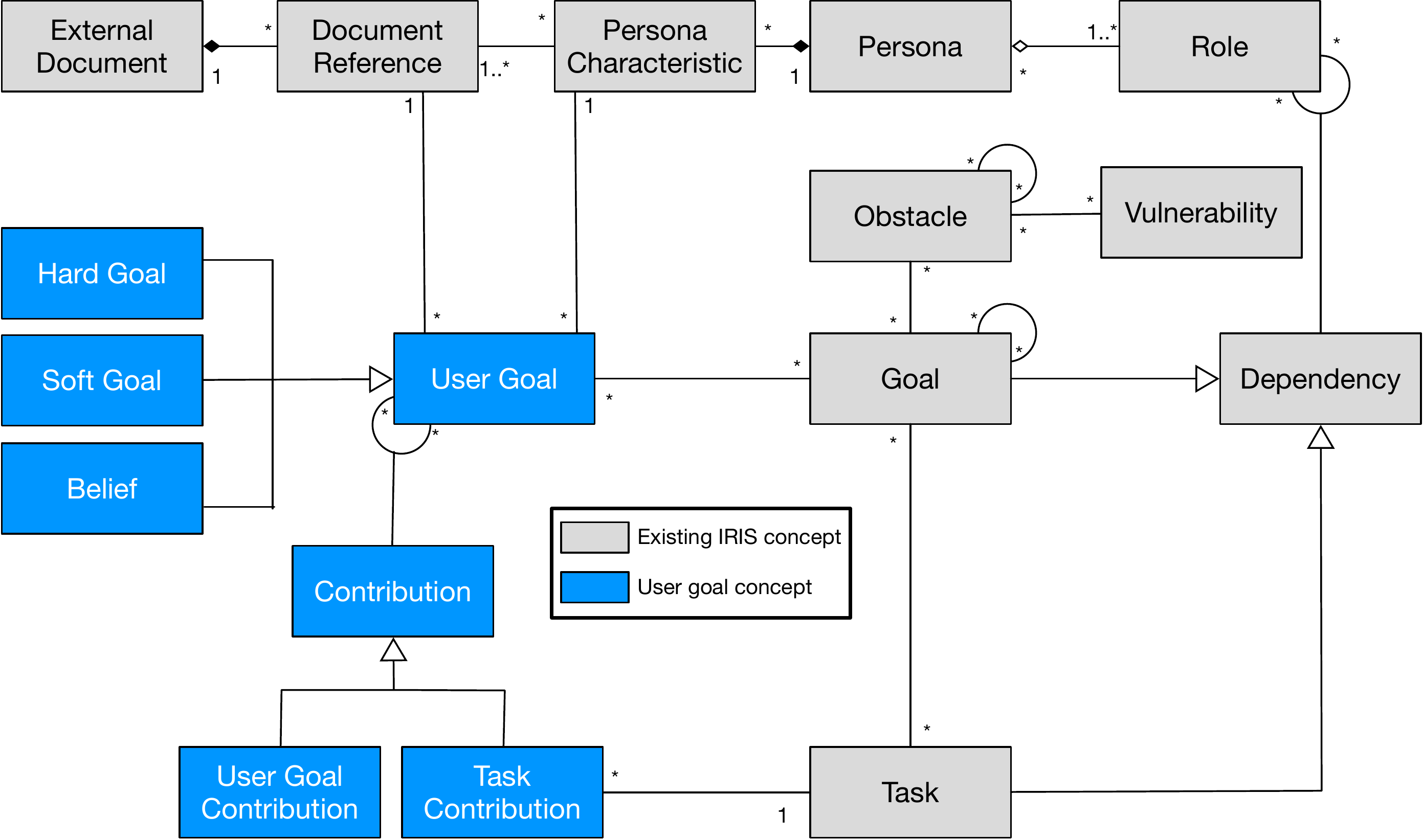}
\caption{UML class diagram of IRIS and user goal concepts}
\label{fig:gug_concepts}
\end{figure}

To reframe a persona as a social goal model, our approach relies on aligning concepts from IRIS with concepts from social goal modelling.  A review of the complete conceptual model, which is described in more detail in  \cite{fail18}, is beyond the scope of this paper.  We do, however, summarise this model concept alignment in Figure \ref{fig:gug_concepts}, which we provide further rationale for in the sub-sections below.

\subsubsection{Personas and Persona Characteristics}
 
Our approach has only a minimal impact on existing IRIS concepts.  We assume personas consist of multiple persona characteristics.  These characteristics are attributes of persona behaviour; they can be considered as arguments for persona behaviour, and are grounded in one or more grounds, warrants, and rebuttal elements.  These elements (document references) are \emph{factoids} that can be drawn from a variety of primary and secondary data sources (external documents) such as interview transcripts, observational notes, and web sites.  Further details on these concepts can be found in \cite{fail18}.

\subsubsection{User Goals}
User goals represent the intentional desires of actors, where actors are personas.  This definition is inline with the definition used for goals by the social goal modelling community, e.g. \cite{yu95}.  In our approach, user goals are factoids expressed intentionally.  Yu et al. \cite{ygmm11} states that intentional properties can only be inferred based on information obtained by indirect means, and that the validity of these attributions can never be certain.  However, a premise of earlier work in the HCI community  \cite{fafl111} is that the qualitative underpinnings of personas \emph{can} be validated, in the same way that qualitative models in general can be validated.  So, although validity can never be certain, our model provides some level of assurance.  Based on the satisfaction levels proposed by Amyot et al. \cite{amgh10}, user goals can be assigned a qualitative satisfaction level associated with a quantitative score; these values are Satisfied (100), Weakly Satisfied (50), Weakly Denied (-50), and Denied (-100).

\subsubsection{Hard/Soft Goals and Beliefs}

Our approach inherits the idea of hard goals, soft goals, and beliefs from \emph{i*}.  Hard goals are goals that can be measurably satisfied, whereas soft goals are goals with less well-defined success criteria that can be satisficed \cite{sim79}.  Beliefs capture facts important to stakeholders \cite{amgh10}; we use these to capture beliefs held by personas.   Beliefs are used irregularly in goal models, and while it has been suggested these are used to capture the rationale of designers during modelling rather than stakeholders \cite{rewe05}, it has also been accepted that further exploration on the semantics of beliefs is needed \cite{yu09}.  The grounding of personas and IRIS' support for KAOS domain properties -- that can capture this form of rationale -- means we need not explicitly incorporate rationale meta-data into visual models.  Therefore, beliefs can be safely used to represent stakeholder beliefs without confusion.  User goals are elicited from persona characteristic elements based on the trust characteristic elicitation process described in \cite{fafl142}, where implied goal, soft goal or belief intentions form the basis of user goals associated with the characteristic and its grounds, warrants and rebuttal elements.  These user goals are expressed as persona intentions.

\subsubsection{Aligning System and User Goals}

As Figure \ref{fig:gug_concepts} shows, IRIS supports the concept of system goal, i.e. prescriptive statements of intent that the system should satisfy throughout the co-operation of its intended roles in a particular environment; this definition is based on the KAOS definition of goal \cite{lams09}.  Obstacles obstructing these goals may be associated with vulnerabilities, thereby connecting a goal view of a system with a risk view.  IRIS also supports dependency modelling of system goals, where a \emph{depender} role depends on \emph{dependee} role for a goal or task \emph{dependum}.   

Until now, IRIS has not incorporated the notion of user goal because, as a methodologically agnostic meta-model, discretion on how to map user goals and expectations to system functionality is left to designers.  However, in the case of a goal dependum, we should be able to capture the need for user goals to be satisfied to satisfy system goals.  Consequently, our approach now adds an explicit traceability link between user goals that personas might have, and KAOS goals that a system needs to satisfy.  This traceability link could be bi-directional, as we do not prescribe the elicitation of one type of goal before the other.  For example, an analyst may capture system goals to satisfy a persona's goals, so may wish to indicate the system goals that address these user goals.  Conversely, in a pre-existing system model, an analyst may wish to examine the implication of system requirements on the value a persona wishes to achieve.  Our approach precludes neither possibility, and facilitates subsequent model validation checks.

\subsection{Modelling user goal contributions}
\label{sect:mgc}

To visualise personas as goal models, our approach extends the \emph{i*} Strategic Rationale model \cite{yu95} in two ways.  First, we align persona characteristic elements with \emph{contribution} links.  Contribution links indicate the desired impact that one system element has on another \cite{amgh10}.  As user goals are part of the broader socio-technical system being modelled, it is reasonable to assume that one user goal can contribute to another.  In our approach, argumentation elements form the basis of means/end contribution links between user goals, i.e. where one user goal is the \emph{means} for another user goal's \emph{end}.  Links are annotated with two additional pieces of information: (i) whether a link is a `means' or an `end' with respect to the characteristic's goal, soft goal or belief, (ii) an optional initial satisfaction level, based on the qualitative values and quantitative scores specified in \cite{amgh10}, i.e. Satisfied (100) Weakly Satisfied (50), Weakly Denied (-50), and Denied (-100); this is analogous to the setting of \emph{strategies} in jUCMNav \cite{amgh10}.  Second, as tasks can have a security impact \cite{elyu09}, completion of a task contributes to one or more user goals.

\begin{algorithm}[H]
  \scriptsize
  \SetKwInOut{Input}{Input}
  \SetKwInOut{Output}{Output}
  \SetAlgoLined
  \Input{$goalName$ - the goal name}
  \KwData{$evaluatedGoals$ - set of previously evaluated goals and their contribution scores, $cts$ - names of tasks contributing to user goal $goalName$, $cgs$ - names of user goals contributing to user goal $goalName$, $linkScore$ - quantitative score for the contribution of user goal $cgName$ to user goal $goalName$, $contScore$ - product of $linkScore$ and the goal contribution score for user goal $cgName$}
  \Output{$score$ - contribution score}

  \SetKwProg{Fn}{Function}{ is}{end}
  \Fn{calculateGoalContribution($goalName$)}{
  
    $score$ $\leftarrow$ initialSatisfactionScore $goalName$;
  
    \If{$score$ = 0}{
      $isObstructed$ $\leftarrow$ systemGoalObstructed $goalName$;
  
      \uIf{$isObstructed$} {
        $score$ $\leftarrow$ -100;
      }
  
      \Else{
        \uIf{$goalName$ $\notin$ domain $evaluatedGoals$}{
    
          $cts$ $\leftarrow$ taskLinks \ $goalName$;
  
          \While{taskName $\leftarrow$ cts} {
  
            $score$ $\leftarrow$ $score$ + taskContributionScore $taskName$;
    
          }
  
          $cgs$ $\leftarrow$ $goalContributions$\ $goalName$;

          \While{$cgName$ $\leftarrow$ $cgs$} {
            $linkScore$ $\leftarrow$ contributionLinkScore\ $goalName$\ $cgName$;   
      
            $cgScore$ $\leftarrow$ calculateGoalContribution\ $cgName$;
      
            $contScore$ $\leftarrow$ $linkScore$ $\times$ $cgScore$ ;
      
            $score$ $\leftarrow$ $score$ + $contScore$; 
          }
    
          $score$ $\leftarrow$ $score$ / 100;
    
          \uIf{$score$ $<$ -100}{
            $score$ $\leftarrow$ -100;      
          }
          \ElseIf{$score$ $>$ 100} {
            $score$ $\leftarrow$ 100;
          }
        
          $evaluatedGoals$ $\leftarrow$ $evaluatedGoals$ $\cup$ \{$goalName$ $\fun$ $score$\};
        }
        \Else{
          $score$ $\leftarrow$ $evaluatedGoals$\ $goalName$;
        }
      }
    }
    \textbf{return} $score$;
  }
  \caption{calculateGoalContribution}
  \label{alg:calcCont}
\end{algorithm}	

Like other goal modelling languages, contributions have a qualitative value corresponding to a quantitative score.  We base these values on those used by GRL: Make (100), SomePositive (50), Help (25), Hurts (-25), SomeNegative (-50), and Break (-100).  Make and Break contributions lead to the satisfaction or denial of user goals respectively; similarly, Help and Hurt contributions help or hinder satisfaction of user goals.  SomePositive and SomeNegative values indicate some indeterminate level of positive or negative contribution that exceeds helping or hindering.

The approach for calculating contributions is similar to Giorgini et al.'s label propagation algorithm \cite{gmns03}.  We implemented a recursive, forward propagation $calculateGoalContribution$ (Algorithm 1) based on the $CalculateContribution$ algorithm described in \cite{amgh10}.  

The setting of an initial satisfaction score (Line 2) based on the previously described satisfaction level is permitted; this can override the calculated goal score from related task and goal contributions.  If the initial satisfaction score has not been overridden and no system goals associated with a user goal have not been obstructed (Lines 4--6), a contribution score is calculated.  To handle goal contribution loops, i.e. where user goal $x$ is a means to goal $y$, which is a means to goal $x$, or situations where the user goal $x$ contributes to several user goals that eventually contribute to user goal $y$, a persistent set of visited goals and their contribution scores, $evaluatedGoals$, is retained.  Propagation occurs if a goal's name is not in this set (Lines 9--26), otherwise the previously retained contribution for that goal is reused (Line 28).  The contribution score is calculated based on the tasks contributing to it (Lines 9--12), and the product of each contributing goal and the contribution link strength (Lines 13--19).  If the score calculated is greater than 100 or less than -100 then the score is normalised to a value within this range (Lines 21--25). 

\subsection{Identifying implicit vulnerabilities}

Our approach for identifying implicit vulnerabilities, which is concerned with dependencies between system rather than user goals, identifies two situations where dependums might not be delivered.  First, if a system goal dependum or its refinements are obstructed and not resolved.  Second, if the dependum or its refinements are linked with denied user goals.  

\begin{algorithm}
  \scriptsize
  \SetKwInOut{Input}{Input}
  \SetKwInOut{Output}{Output}
  \SetAlgoLined

  \Input{$g$ - the goal name}
  \KwData{$ugs$ - names of user goals linked to system goal $g$, $goals$ - names of system goals refinements of $g$, $obs$ - names of obstacles obstructing system goal $g$}
  \Output{$isObstructed$ - indicates if goal $g$ is obstructed}
		
  \SetKwProg{Fn}{Function}{ is}{end}
  \Fn{isGoalObstructed($g$)}{
  
    $isObstructed$ $\leftarrow$ false;
    
    $ugs$ $\leftarrow$ linkedUserGoals $g$;
   
    \While{ug $\leftarrow$ ugs}{
      $score$ $\leftarrow$ calculateGoalContribution\ $ug$\ [];
    
      \If{$score$ $<$ 0}{
        $isObstructed$ $\leftarrow$ true;
      
        break;
      }
    }
  
    \If{isObstructed = false}{
  
      $goals$ $\leftarrow$ refinedGoals $g$;
			
      \eIf{$goals$ = $\emptyset$}{
        $obs$ $\leftarrow$ obstructingGoals\ $g$;
			  
        \eIf{$obs$ $\neq$ $\emptyset$}{
          $isObstructed$ $\leftarrow$ true;	
        }{
          \While{$o$ $\leftarrow$ $obs$}{
            $isObstructed$ $\leftarrow$ isObstacleObstructed\ o;
				
          }
		      
        }
			  			  
      }{
  
        \While{$g$ $\leftarrow$ $goals$}{
	      $isObstructed$ $\leftarrow$ isGoalObstructed $g$;
	    }
      }
    }
    \textbf{return} $isObstructed$;
  }
\caption{isGoalObstructed check}
\label{alg:igo}
\end{algorithm}	
	
Algorithm \ref{alg:igo} specifies how the presence of such implicit vulnerabilities might be identified within a typical recursive system goal satisfaction algorithm.  The algorithm returns a value of true if the system goal $g$ is obstructed. 

The algorithm navigates the operationalising tree-based KAOS goal refinements (Lines 11--27) to determine if there are \emph{obstruct} associations between refined goals and obstacles, and these obstacles have not been resolved, i.e. there are no \emph{resolve} relationships between obstacles and goals which address them.  However, this check can be shortcut should a linked user goal associated with system goal $g$ be denied, i.e. has a score less than 0. (Lines 3--10).  Should this check not be shortcut then the $isObstacleObstructed$ algorithm (Line 19) determines whether a goal is obstructed.  This algorithm returns a value of true should one or more of the following conditions hold: (i) the obstacle or one of its obstacle refinements are not resolved by a [mitigating] system goal, (ii) an obstacle or one of its obstacle refinements are resolved, but the resolved goal has one or more linked user goals which are denied.  The $isObstacleObstructed$ algorithm is formally specified in \cite{fsss20}. 

Vulnerabilities within IRIS are defined as system weaknesses \cite{fail18}, but an implicit vulnerability may not always be a system weakness.  It may indicate some inconsistency between what system roles and humans fulfilling might want and need, or -- as suggested by \cite{paem11} -- some level of human fallibility resulting from roles that participate in too many dependencies as a depender.  However, implicit vulnerabilities can help make sense of different system models and, in doing so, provide rationale for vulnerabilities feeding into risk models.

\subsection{Tool-support}

To show how this approach might be implemented in Requirements Management tools more generally, we incorporated a new model type and supporting tools into CAIRIS release 2.3.6.

We tool-supported the additional concepts and algorithms by introducing a \emph{User goal} visual model.  This is based on the visual semantics of GRL, where a rounded box represents a hard goal, a polygon with rounded corners represents a soft goal, an ellipse represents a belief, and a dashed rectangle models the actor boundary.  In this model, actors are represented by personas.  Further drawing from the semantics used by GRL and jUCMNav, these nodes are coloured from dark green to dark red corresponding with satisfaction values of Satisfied (100) and Denied (-100); nodes with a value of None (0) are coloured yellow.  

User goal models are generated automatically by CAIRIS using the same pipeline process used to visualise other CAIRIS models.  A declarative model of graph edges is generated by CAIRIS; this is processed and annotated by graphviz \cite{graphviz} before being subsequently rendered as SVG.  This annotation stage includes applying Algorithm 1 to user goal nodes to determine its score, and subsequent colour.  The CAIRIS model generation process is described in more detail in \cite{fail18}.  The algorithms described were incorporated into a \emph{Implied vulnerability} model validation check, which is applied to all KAOS goal dependency relationships in a CAIRIS model.  CAIRIS model validation checks are implemented internally within the relational database used by a CAIRIS model as SQL stored procedures. 

\begin{figure*}[h!]
\centering
\includegraphics[scale=0.32]{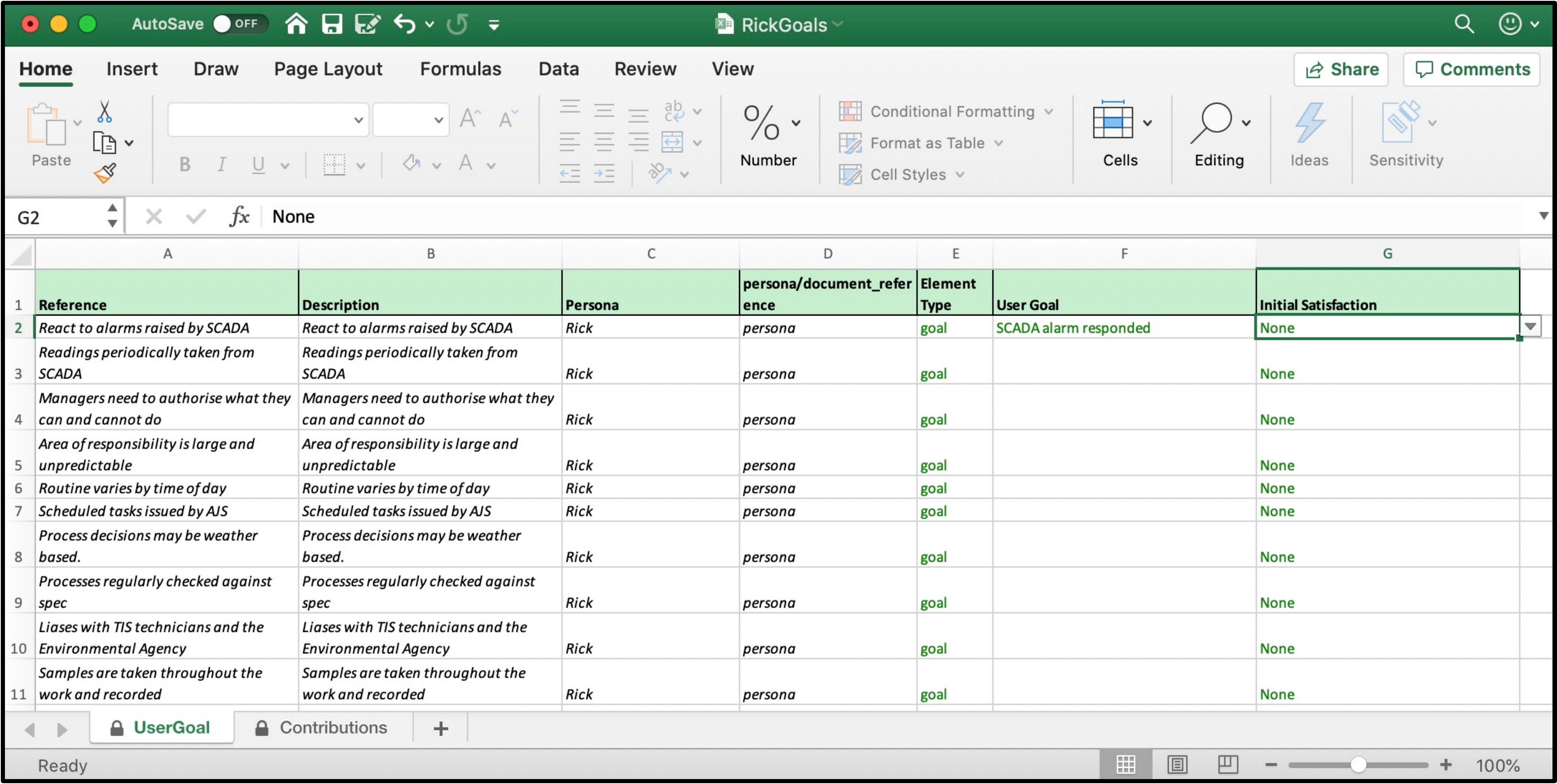}
\caption{Generated Excel workbook for entering user goals and contributions}
\label{fig:excel}
\end{figure*}

As shown in Figure \ref{fig:excel}, we also extended CAIRIS to generate Excel workbooks for capturing user goals and contribution links.  Such workbooks are useful for analysts wishing to contribute to user goal modelling via more familiar office automation tools.

The generated Excel workbook contains UserGoal and UserContribution spreadsheets, where edited cells for both are coloured green.  The UserGoal worksheet is pre-populated with read-only data on the persona characteristic or document reference name, its description, the persona it is associated with, and an indicator to whether the reference corresponds to a persona [characteristic] or document reference. When completing the worksheet, analysts should indicate the intentional elements associated with the persona characteristics or document references providing their grounds, warrants, or rebuttals. Analysts should also indicate the element type (goal, softgoal, or belief), and the initial satisfaction level using the dropdown lists provided.  The source and destination cells in the ContributionsSheet are pre-populated once user goals have been added in the UserGoal sheet, so only the means/end and contribution links need to be set. 

We further extended CAIRIS to allow the contents of these workbooks to be imported into a pre-existing CAIRIS model.

\section{Case Study}
\label{sect:results}

\subsection{ACME Water Security Policy}

We evaluated our approach by using it to identify implicit vulnerabilities associated with the security policy of \emph{ACME Water}: an anonymised UK water company responsible for providing clean and waste water services to several million people in a particular UK region.  The infrastructure needed to support such a large customer base was substantial, amounting to over 60 water treatment works, 800 waste water treatment works, 550 service reservoirs,  27,000 km of water mains, 19,000 km of sewer networks, with over 1,900 pumping stations, and 3,200 combined sewer outflows.  This policy was modelled as a KAOS goal model where each system goal represented a policy goal.  

Four in-situ interviews were held with 6 plant operators, SCADA engineers and plant operation managers at two clean water and two waste water treatment plants.  These interviews were recorded, and the transcripts analysed using Grounded Theory.  The results of this analysis are a qualitative model of plant operations security perceptions.  Using the persona case technique \cite{fafl111}, we analysed the Grounded Theory model to derive a single persona of a water-treatment plant operator, Rick, incorporating 32 persona characteristics, and backed up by 82 argumentation elements (grounds, warrants, or rebuttals). 
 
\begin{figure*}[h!]
\centering
\includegraphics[scale=0.28]{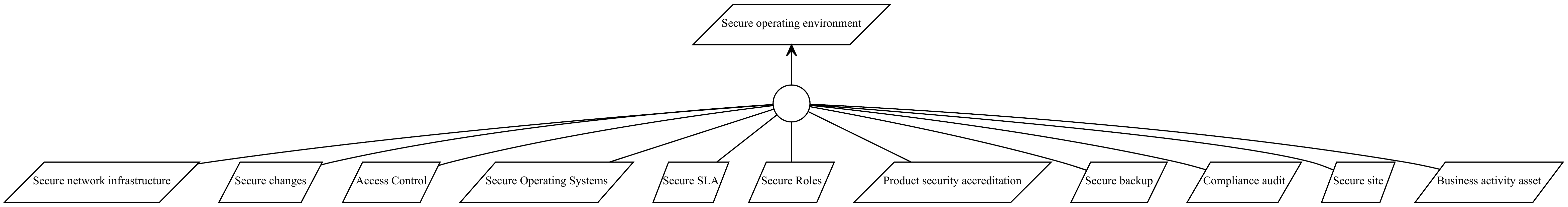}
\caption{High-level ACME Water security policy goals}
\label{fig:pgs}
\end{figure*}

The security policy goals were created by analysing existing documentation about ACME's existing information security policy and agreeing the scope of the policy to be modelled with ACME's IT security manager.  Existing policy documentation was analysed to elicit and specify a KAOS goal model of 82 policy goals, with a single high level goal (\emph{Secure operating environment}) and, as shown in Figure \ref{fig:pgs}, 11 refined sub-goals representing the different policy areas.  These goals and other security and usability elements of the operating environment were specified in a CAIRIS model; these included 2 personas, 11 roles, 21 obstacles, 9 vulnerabilities, 5 tasks, and 6 role-goal-role dependencies. \footnote{The case study CAIRIS model is available from \url{https://doi.org/10.5281/zenodo.3979236}}
    
\subsection{User goal model creation}

To generate a user goal model based on Rick, we initially derived 104 user goals and beliefs from both the persona characteristics and argumentation elements, and 165 contribution links.  The first two authors then reviewed the model to de-duplicate synonymous user goals.  For example, a \emph{Site protected} user goal was associated with a \emph{Copper theft} document reference, as the intention implied was that the site needed to be protected from this threat.  However, we identified a \emph{Site secured} user goal associated with \emph{Physical and login security} document reference.  As a result, we deleted the former user goal, and contribution linked its user goals to \emph{Site secure}.  In parallel with the de-duplication of user goals, we also added additional contribution links between user goals based on our understanding of the persona and his intentions, where these contribution links cross-cut persona characteristics.  For example, on reviewing the persona characteristics and their underpinning data, we noted that the \emph{Thieves ignore impact} user goal, which was associated with the \emph{Thieves do not care about their impact} characteristic, helped foster the belief that \emph{Personal safety is a hygiene factor}; this belief was associated with the \emph{Personal safety is an infosec hygiene factor} persona characteristic.  Following this analysis, the final model resulted in 93 user goals and beliefs, and 205 contribution links.

\begin{landscape}
\begin{figure}
\centering
\includegraphics[scale=0.52]{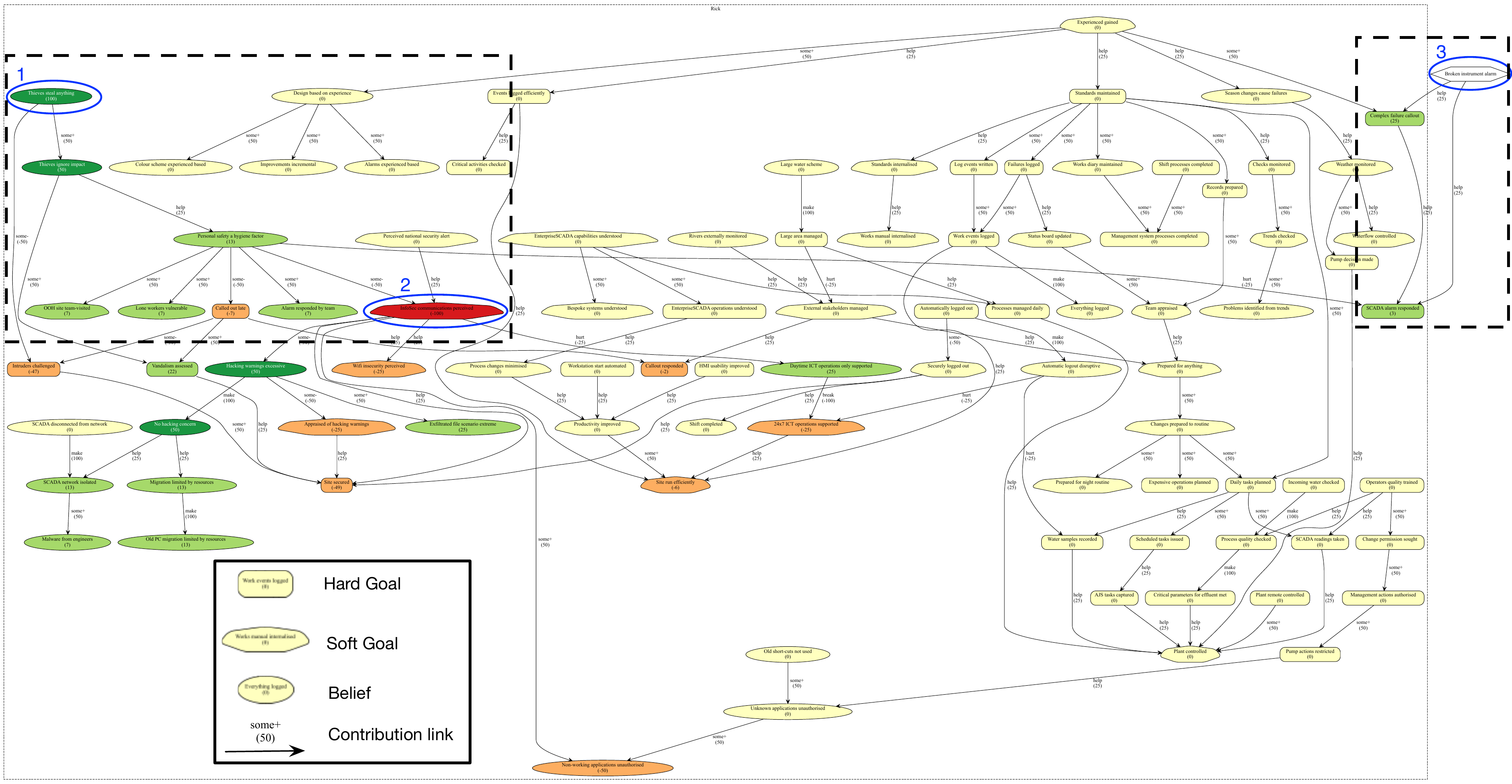}
\caption{Annotated CAIRIS User Goal model based on Rick persona}
\label{fig:fgm}
\end{figure}
\end{landscape}

Figure \ref{fig:fgm} shows the goal model generated by CAIRIS for Rick.  

\subsection{ICT awareness implicit vulnerabilities}

\begin{figure*}[h!]
\centering
\includegraphics[scale=0.29]{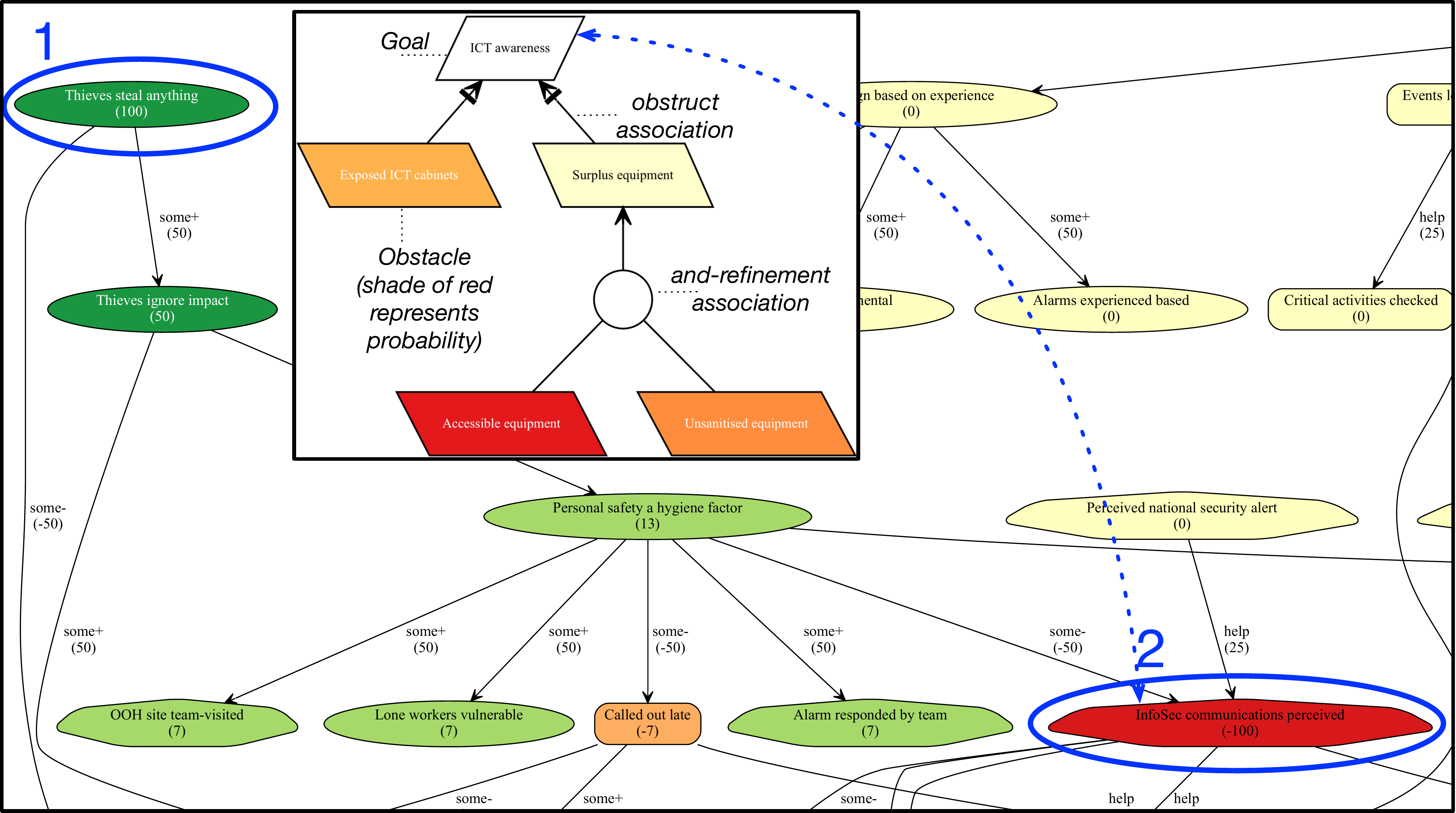}
\caption{Alignment between \emph{ICT awareness} system goal in KAOS goal model (inset) and \emph{InfoSec communications perceived} user goal in user goal model}
\label{fig:annot1}
\end{figure*}

From Figure \ref{fig:annot1}, we identified a link between the \emph{InfoSec communications perceived} user goal (annotated as 2) and the \emph{ICT awareness} system goal, which is a refinement of the high-level \emph{Secure Site} system goal.

The \emph{ICT awareness} system goal indicates ICT partners should know how to maintain equipment hosted in the secure areas and, as Figure \ref{fig:annot1} (inset) shows, this system goal is already obstructed due to exposed and surplus equipment which should not be present.  Unfortunately, as Figure \ref{fig:annot1} also indicates, the related user goal is also denied.  The negative impact affects not only the perception of site security, but also the perception the site is run efficiently; this corroborates the obstacles found to be present in the system goal model.  To reinforce this, the belief \emph{Thieves steal anything} (annotated as 1) was set to satisfied, which weakly denied \emph{InfoSec communications perceived}, further validating negative perception.  This highlighted the need for a new dependency where an IT security manager depends on ICT partner to achieve the \emph{ICT awareness} goal.

The limited security awareness means operators fail to see the connection between misunderstanding authorisation, and wifi insecurity and site security, due to their belief than an air-gap exists between wireless networks and industrial control systems.  Access controls on pump actions further supports the belief that unknown applications are unauthorised.  To explore this further, we associated the \emph{Pump action restricted} user goal with the \emph{Access Control} system goal, and added a dependency to indicate that plant operators depend on Information Security managers for this goal.  CAIRIS subsequently flagged a model validation warning because a refined goal\emph{Vendor passwords} was obstructed, due to evidence that vendors were using easily guessed default passwords for certain critical components.  

\subsection{Validating vulnerabilities with implicit vulnerabilities}

As indicated in Figure \ref{fig:gug_concepts}, obstacles can be associated with vulnerabilities to capture the rationale for including vulnerabilities in subsequent risk analysis activities.  In the ACME Water model, an \emph{Exposed ICT Cabinets} obstacle was already associated with an \emph{Exposed cabinet} vulnerability, but - given how divisive resolving obstacles might be due to the architectural implications of their resolution -- we wanted to see if the user goal model of Rick provided a human rationale for the obstacle's presence.
   
Information Security Managers depend on Plant operators for a related \emph{Industrialised secure cabinet} system goal to ensure control systems are kept in secure cabinets.  On reviewing the user goal model and the tasks in the ACME Water model, we noted that no-one was explicitly required to check these cabinets; instead, ACME Water trusted Rick to do this while discharging other duties.

\begin{figure*}[h!]
\centering
\includegraphics[scale=0.4]{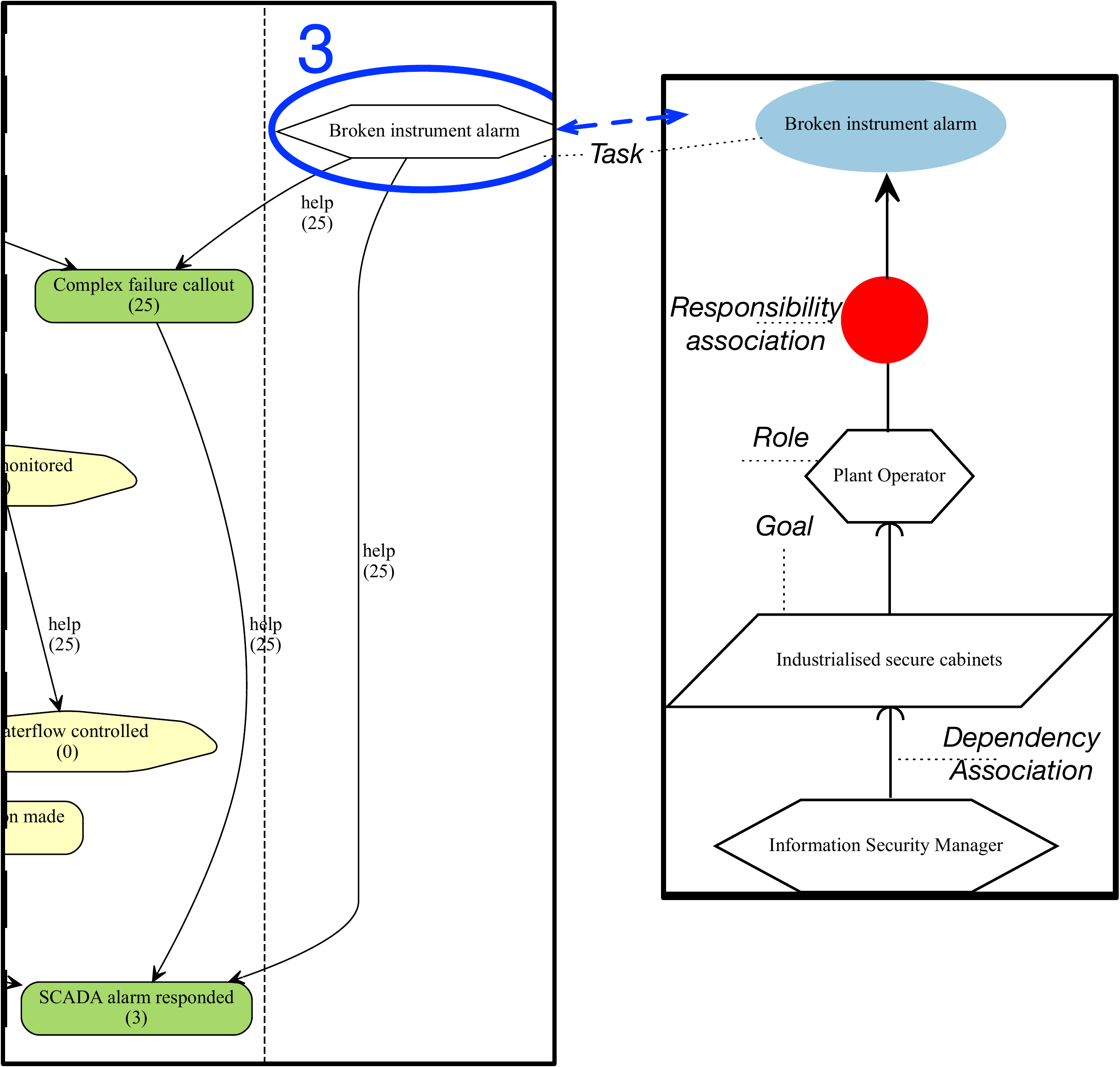}
\caption{Contribution of \emph{Broken Instrument alarm} task to user goals (left) and related responsibility and dependency associations (right)}
\label{fig:annot2}
\end{figure*}

As Figure \ref{fig:annot2} shows, as part of a pre-existing \emph{Broken Instrument alarm} task (annotated as 3), we introduced help contribution links to \emph{Complex failure callout} and \emph{SCADA alarm responded} because Rick completes the task to satisfy these user goals.  The task entails Rick being away from the safety of the control room to respond to equipment alarms from these cabinets.  Should these alarms fire out of hours, the model shows that Rick might feel uneasy, particularly if he thinks the alarm indicates intruders are stealing equipment.  The potential for Rick to skip the steps necessary to check these cabinets was corroborated in the user goal model due the \emph{SCADA alarm responded} being very weakly satisfied.
  
\section{Discussion and Limitations}
\label{sect:discussion}

While important for validating requirements, traceability is a weakness of languages like \emph{i*} due to lack of guidelines for working with complementary models \cite{paem11}.  Our approach addresses this traceability problem by drawing user goal relationships from the qualitative data analysis underpinning personas.  However, a limitation of our approach is the restricted expressiveness of the generated user goal models, particularly the lack of support for \emph{strategic dependencies} between user goals.  Supporting dependencies between user goals may appear trivial from a modelling perspective, but retaining traceability would necessitate changes to how the qualitative data grounding personas is elicited and analysed to ensure both personas and their collaborative aspects are encapsulated.  Approaches for creating such personas already exist, e.g. \cite{mawh11}, and could provide a grounding for subsequent modelling of user goal dependencies.

Another limitation of our work is that our case study considers only a single persona.  However, our initial results developing and evaluating the changes to CAIRIS indicate that user goal models place little additional performance burden to model validation checks.  Because CAIRIS can incrementally import models that overlay existing models, it is possible to incrementally add personas to a baseline system to explore the impact of different personas interacting with each other.  Based on the process and performance of the tool-support, we believe our approach scales to multiple personas too, but a more thorough performance evaluation will be the subject of future work.

\section{Conclusion}
\label{sect:conclusion}

This paper presented an approach for reframing personas as social goal models and, in doing so, using both the reframed and related models to find implicit vulnerabilities.  As a result, we have made two contributions addressing our research questions in Section \ref{sect:introduction}.  First, we addressed RQ1 by demonstrating how the user research used to construct personas can be leveraged to partially automate construction of social goal models.  Such user goals could be elicited either while constructing personas, or afterwards - in which case the process of constructing the user goal models helps further validate the personas and the data upon which they are based.   Second, we addressed RQ2 by illustrating how minimal contributions to existing tool-support facilitate automation for both the identification of implicit vulnerabilities from user goal models, and the validation of existing system goal obstructions based on user goals and user goal contributions.  Our intention is not to replace traditional RE approaches to system and social goal modelling, but to show how applying them in a different way can identify and confirm potential security problems that might have otherwise remained hidden.

Future work will further examine persona characteristics and goal and task attributes to evaluate fitness between persona and actors in goal models.  For example, some goals might require long-term attention span while others require different social skills.  The user model associated with these attributes will be then used to simulate how different personas interact, and whether this leads to insecurity.  We will also investigate collaborative information gathering techniques to capture goal models and their personas, e.g. through an interactive algorithm driven by representative users providing satisfaction and denial weights, and propagation options. 

\bibliographystyle{spmpsci}
\bibliography{manuscript}
\end{document}